\documentclass[aps,twocolumn,superscriptaddress,showpacs]{revtex4}
\usepackage{amsmath}
\usepackage{graphicx}
\usepackage{xcolor}
\begin{document}

\title{Controlling Quasibound States in 1D Continuum Through Electromagnetic
Induced Transparency Mechanism}
\author{Z. R. Gong}
\affiliation{Institute of Theoretical Physics, The Chinese Academy of Sciences, Beijing,
100080, China}
\author{H. Ian}
\affiliation{Institute of Theoretical Physics, The Chinese Academy of Sciences, Beijing,
100080, China}
\author{Lan Zhou}
\affiliation{Department of Physics, Hunan Normal University, Changsha 410081, China}
\affiliation{Frontier Research System, The Institute of Physical and Chemical Research
(RIKEN), Wako-shi 351-0198, Japan}
\author{C. P. Sun}
\affiliation{Institute of Theoretical Physics, The Chinese Academy of Sciences, Beijing,
100080, China}
\date{\today}

\begin{abstract}
We study the coherent scattering process of a single photon confined in an
one-dimensional (1D) coupled cavity-array, where a $\Lambda$-type
three-level atom is placed inside one of the cavities in the array and
behaves as a functional quantum node (FQN). We show that, through the
electromagnetic induced transparency (EIT) mechanism, the $\Lambda$-type FQN
bears complete control over the reflection and transmission of the incident
photon along the cavity-array. We also demonstrate the emergence of a
quasibound state of the single photon inside a secondary cavity constructed
by two distant FQN's as two end mirrors, from which we are motivated to
design an all-optical single photon storage device of quantum coherence.
\end{abstract}

\pacs{42.50.Gy, 32.80.Qk, 03.65.Nk, 42.70.Qs}
\maketitle

\section{\label{sec:introduction}INTRODUCTION}

In recent years, many efforts have been exerted to implement all-optical
quantum devices~\cite{allopt-1,allopt-2,allopt-3,allopt-4} that can
coherently control photons through photons~\cite%
{p2p-1,p2p-2,p2p-3,p2p-4,p2p-5}. This application-oriented pursuit requires
the possible existence of a strong and controllable photon-photon
interaction. From the view of modern physics, photons do not couple to each
other directly through fundamental electromagnetic interactions; however,
people recognized that two photons interact indirectly via nonlinear media~%
\cite{Kerr-1,Kerr-2,Kerr-3,Kerr-4,Kerr-5}. Such nonlinear interactions are
usually obtained through high-order perturbation theories and hence cannot
be arbitrarily manipulated according to one's will. Nevertheless, inside
some artificial medium, the transport of photons can be well controlled by
an additional intervening classical field~\cite{allopt-1}.

In this paper, we revisit this problem of photon transportation under a
coherent architecture, based on the theoretical approach we have developed
in Ref.~\cite{zhou}. We propose that a coupled cavity-array, regarded as an
one-dimensional (1D) continuum, provides a transport channel to an incident
single photon. Placed inside one of the cavities, a $\Lambda$-type
three-level atom can either grant or block the path of the single photon by
the atom's electromagnetic induced transparency (EIT) effect. This atom can
essentially be regarded as a functional quantum node (FQN). In fact, a
similar mechanism has been used to build the so-called single-photon
transistor, analogous to an electronic transistor in which an atom plays the
role of the gate through its absorption and emission of photons to-and-fro a
channel~\cite{allopt-2}.

In comparison with the design of the {}``single-photon transistor'', in
which the photon travels continuously through a fiber waveguide, our
cavity-array lets the photon travel discretely through the channel by
locally creating or annihilating a photon between its cavities. We hence
consider the coherent scattering process of the photon with the
forementioned FQN in the discrete coordinate representation. Our approach is
then a generalization of the 1D process where a photon is scattered by the $%
\delta$-potential set up by a FQN through its EIT effect into the discrete
space. Such an approach raises a richer spectrum structure indicated by its
nonlinear dispersion relation resulting from the tight-binding inter-cavity
coupling, as opposed to the usual linear dispersion relations. Because of
the unusual dispersion relation, we develop a new approach for the transport
of the single photon, which is different from the effective field approach~%
\cite{fan-1} normally adopted; yet the high energy limit of our setup can
cover the main results of the theoretical approaches of similar
single-photon transistor designs~\cite{fan-1,fan-2,fan-3}.

We generalize the discrete scattering method recently proposed~\cite{zhou}
and calculate the reflection and the transmission coefficients of the
single-photon transport as functions of both the Rabi frequency and the
level spacing between the excited state and the metastable state of the FQN.
The reflection and the transmission spectra of the photon are depicted by
general lineshapes whose ranges cover both the high-energy end at the
Breit-Wigner limit and the low-energy end at the Fano-Feshbach limit. As
emphasized, it is the classical field that control the system to reach such
high energy and low energy limits. Our derivations show that the $\Lambda$%
-type FQN can behave as a perfect mirror to totally reflect the incident
photon when the Rabi frequency matches the frequency of the controlling
classical field. The behavior of the FQN has suggested a basic mechanism to
implement all-optical control for single-photon transports and provides the
ground for a photon storage device through controllable quasibound states,
which are defined in Ref.~\cite{bound-1,bound-2}, inside a secondary cavity
braced by two distant FQN's.

The rest of the paper is organized as follows. In Sec.~\ref{sec:setup}, we
present the model Hamiltonian for a single photon scattered by a $\Lambda$%
-type FQN. In Sec.~\ref{sec:scattering}, we derive the scattering equation
for the transport of the single photon and demonstrate the somewhat
equivalent role played by a $\Lambda$-type FQN and two two-level FQN's at
some particular positions. In Sec.~\ref{sec:reflection}, the reflection and
the transmission coefficients are derived to find the conditions for perfect
reflection and transmission; the spectrum lineshapes at high and low energy
limits are also calculated. In Sec.~\ref{sec:quasibound}, we illustrate the
mechanism we design for photon storage. The conclusion is given in Sec.~\ref%
{sec:conclusion}.

\section{\label{sec:setup}MODEL SETUP}

We consider an 1D coupled cavity array, in which the transport of a single
photon is described by a bosonic tight binding model. A $\Lambda$-type
three-level atom, whose ground state, metastable state and excited state are
denoted as $\left\vert g\right\rangle $, $\left\vert a\right\rangle $ and $%
\left\vert e\right\rangle ,$ respectively, is placed in one of the cavities.
A strong classical field with frequency $\omega_{\mathrm{C}}$ matches to the
$\left\vert e\right\rangle \rightarrow\left\vert a\right\rangle $
transition, while the photon, considered a weak quantum radiation field and
traversing in the cavity array with frequency $\omega_{\mathrm{P}}$, matches
to the $\left\vert e\right\rangle \rightarrow\left\vert g\right\rangle $
transition. The classical field as the control and the quantum field as the
probe dress the three-level atom into an EIT medium. The system and the
detailed structure of the EIT medium are shown in Fig.~\ref{fig:model}(a)
and Fig.~\ref{fig:model}(b), respectively.

The total Hamiltonian
\begin{equation}
H=H_{\mathrm{p}}+H_{\mathrm{a}}+H_{\mathrm{c}}  \label{2-1}
\end{equation}%
contains three parts, describing in order: the propagation of the photon,
the free $\Lambda $-type three-level atom, and the controlling couplings

\begin{subequations}
\begin{align}
H_{\mathrm{p}}& =\underset{j}{\sum }[\omega b_{j}^{\dag }b_{j}-t(b_{j}^{\dag
}b_{j+1}+h.c.)],  \label{2-2-1} \\
H_{\mathrm{a}}& =\omega _{\mathrm{e}}\left\vert e\right\rangle \left\langle
e\right\vert +\omega _{\mathrm{a}}\left\vert a\right\rangle \left\langle
a\right\vert ,  \label{2-2-2} \\
H_{\mathrm{c}}& =\Omega (e^{-i\omega _{\mathrm{C}}t}\left\vert
a\right\rangle \left\langle e\right\vert +h.c.)+g(b_{0}^{\dag }\left\vert
g\right\rangle \left\langle e\right\vert +h.c.),  \label{2-2-3}
\end{align}%
where $b_{j}^{\dag }$ is the creation operator of photon at site $j$ with $%
\omega $ the cavity field frequency and $t$ the hopping coefficient; $\omega
_{\mathrm{e}}$ and $\omega _{\mathrm{a}}$ are the energy level spacings of
the metastable state and the excited state, respectively; $\Omega $ is the
Rabi frequency due to the control field and $g$ the coupling constant to the
probe field.

%
\begin{figure}[ptb]
\begin{centering}
\includegraphics[bb=30bp 382bp 547bp 749bp,clip,width=3in]{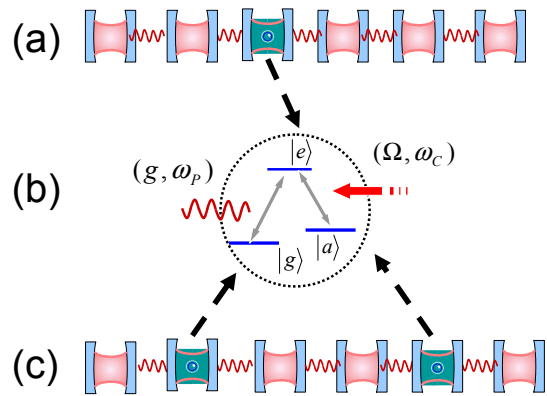}
\par\end{centering}
\caption{(Color online) Schematic of the model setup. (a) shows the
1D coupled cavity-array with one deposited $\Lambda$-type atom. (b)
shows the internal structure of the $\Lambda$-type atom along with
its coupling light fields, where the red curvy line denotes the
quantum light field propagating in the cavity-array and the red
arrow denotes the classical light field. (c) shows the cavity-array
with two $\Lambda$-type atoms deposited at different sites, which
forms a secondary cavity.} \label{fig:model}
\end{figure}


We first consider the scattering process in a {}\textquotedblleft
rotating\textquotedblright frame of reference, which is defined by a unitary
transformation $U=\exp (-i\omega _{\mathrm{C}}\left\vert a\right\rangle
\left\langle a\right\vert t).$Then the original Hamiltonian is changed into
a time independent one as
\end{subequations}
\begin{align}
H^{R}& =U^{\dag }HU-iU^{\dag }\partial _{t}U  \notag \\
& =H_{\mathrm{p}}+H_{\mathrm{a}}^{\prime }+H_{\mathrm{c}}^{\prime },
\label{2--3}
\end{align}%
where, corresponding to renomalized $\Lambda $-type atom and controlling
interaction, respectively,
\begin{subequations}
\begin{align}
H_{\mathrm{a}}^{\prime }& =\omega _{\mathrm{e}}\left\vert e\right\rangle
\left\langle e\right\vert +\Delta \left\vert a\right\rangle \left\langle
a\right\vert ,  \label{2-4-1} \\
H_{\mathrm{c}}^{\prime }& =\Omega (\left\vert a\right\rangle \left\langle
e\right\vert +h.c.)+g(b_{0}^{\dag }\left\vert g\right\rangle \left\langle
e\right\vert +h.c.).  \label{2-4-2}
\end{align}
where $\Delta =\omega _{\mathrm{a}}-\omega _{\mathrm{C}}$ is frequency
detuning between the metastable state and the classical light field or the
model's control parameter in the EIT mechanism.

Our setup can be physically implemented in two ways. One is to place
artificial $\Lambda$-type atoms in the 1D defect cavities of a 2D optical
crystals {[}cite:greentree]. The other is to use an 1D superconducting
transmission line of resonators controlled by a three-level Josephson
junction. Both the optical crystal line-cavity and the superconducting
transmission-line resonators provide the 1D continuum for coherent transport
of photons. The controller parts are implemented by an external classical
light field in the case of a $\Lambda$-atom and an external magnetic field
in the case of an artificial Josephson three-level atom.

\section{\label{sec:scattering}DISCRETE SCATTERING BY THE THREE-LEVEL FQN}

To consider the 1D scattering problem for the above model, we divide $H^{%
\mathrm{R}}=H_{0}+H_{\mathrm{I}}$ into two parts where $H_{0}=H_{\mathrm{p}%
}+H_{\mathrm{a}}^{\prime}$ is the free energy part of the cavity-array and
the three-level atom, and $H_{\mathrm{I}}=H_{\mathrm{c}}^{\prime}$ is the
controlling interaction. Through controlling the Rabi frequency and the
cavity mode frequency of the classical field, we can adjust the reflection
and the transmission of light through the atom and thus manipulate the
propagation of the single photon in the 1D continuum.

The single photon process defines a conversation rule of total occupation
number (photons in the cavities plus the excitations of the atom). Written
in the tensor product space, this occupation number operator reads
\end{subequations}
\begin{equation}
\hat{N}=\sum_{j}\left\vert 1_{j},g\right\rangle \left\langle
1_{j},g\right\vert +\left\vert 0,a\right\rangle \left\langle 0,a\right\vert
+\left\vert 0,e\right\rangle \left\langle 0,e\right\vert =1,  \label{3-2}
\end{equation}
which commutes with the total Hamiltonian. In the above identity, $%
\left\vert 1_{j},g\right\rangle $ represents the state in which one photon
occupies the site $j$ while all other sites $i$ with $i\neq j$ have no
photon and the atom is at ground state; $\left\vert 0,e\right\rangle $ and $%
\left\vert 0,a\right\rangle $ represent the states in which no photon exists
in the cavity-array while the atom is promoted to the excited state and the
metastable state, respectively. We hence find an invariant subspace spanned
by the stationary eigenvectors
\begin{equation}
\left\vert E\right\rangle =\underset{j}{\sum}u(j)\left\vert
1_{j},g\right\rangle +u_{\mathrm{a}}\left\vert 0,a\right\rangle +u_{\mathrm{e%
}}\left\vert 0,e\right\rangle .  \label{3-3}
\end{equation}
where $u(j)$, $u_{a}$, $u_{b}$ denotes the probability amplitudes for each
state accordingly.

The eigen-equation $H\left\vert E\right\rangle =E\left\vert E\right\rangle $
results in the system of equations
\begin{subequations}
\begin{align}
(E-\omega)u(j) & =-tu(j+1)-tu(j-1)+gu_{\mathrm{e}}\delta_{j,0},
\label{3-4-1} \\
(E-\omega_{\mathrm{e}})u_{\mathrm{e}} & =gu(0)+\Omega u_{\mathrm{a}},
\label{3-4-2} \\
(E-\Delta)u_{\mathrm{a}} & =\Omega u_{\mathrm{e}},  \label{3-4-3}
\end{align}
about the probability amplitudes. By eliminating the amplitudes for the
atom's excited state and metastable state, we obtain the discrete scattering
equation for the amplitude of the single photon
\end{subequations}
\begin{equation}
[E-\omega-V(j)]u(j)=-tu(j+1)-tu(j-1),  \label{3-5}
\end{equation}
where
\begin{equation}
V(j)=\frac{g^{2}(E-\Delta)}{(E-\omega_{\mathrm{e}})(E-\Delta)-\Omega^{2}}%
\delta_{j,0}=V\delta_{j,0}  \label{3-6}
\end{equation}
is the $\delta$-type effective potential determined by the internal
structure of the $\Lambda$-type atom and $V$ indicates the magnitude of the
potential.

It should be pointed out that the effective potential $V(j)$ is actually
dependent on the eigenenergy $E$. Or inversely the energy $E$ of the
incident photon indirectly determine the magnitude of the effective
potential and can thus render the effective potential smoothly from a
barrier to a well following its variation. Although this scattering
potential is not energy-independent, we can still apply the time-independent
scattering theory in the coordinate space for some certain energy of the
incident photon.

\begin{figure}[tbp]
\begin{centering}
\includegraphics[bb=100bp 345bp 435bp 560bp,clip,width=3in]{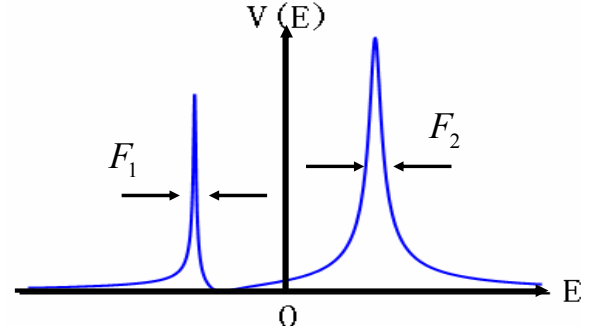}
\par\end{centering}
\caption{Schematic diagram of the potential, where we apply two decay rates
in order to make the values at resonant frequencies finite.}
\label{fig:potential}
\end{figure}

We are interested in the conditions when the total reflection or the total
transmission of the photon controlled by an external classical field occur.
Under such circumstances, the atom behaves like a single photon switch. When
we apply a classical field with a matching frequency, the EIT effect occurs,
there is no photon transport in and out of the cavity-array. The physical
property of the photon transport is totally contained in $V(j).$ Its
magnitude $V$ can be rewritten as%
\begin{equation}
V=g^{2}\left(\frac{A}{E-\omega_{+}}+\frac{B}{E-\omega_{-}}\right),
\label{3-7}
\end{equation}
where the two peaks of maximum magnitude are defined by the two resonant
frequencies
\begin{equation}
\omega_{\pm}=\frac{\omega_{\mathrm{e}}+\Delta}{2}\pm\mu,  \label{3-8}
\end{equation}
and the corresponding amplitudes
\begin{subequations}
\begin{align}
A & =\frac{1}{2}(1+\nu),  \label{3-9-1} \\
B & =\frac{1}{2}(1-\nu).  \label{3-9-2}
\end{align}
In the above identities,
\end{subequations}
\begin{equation}
\mu=\sqrt{\Omega^{2}+\left(\frac{\omega_{\mathrm{e}}-\Delta}{2}\right)^{2}}
\label{3-10}
\end{equation}
denotes the energy splitting between the two resonant frequencies and
\begin{equation}
\nu=\frac{\omega_{\mathrm{e}}-\Delta}{2\mu}.  \label{3-11}
\end{equation}
the amplitude difference between the two peaks. The absolute value of the
potential is plotted in Fig. \ref{fig:potential}, which explicitly shows the
two peaks at two different values of energy $E$. In order to make the
potential finite, we have phenomenologically introduced two additional decay
rates $\Gamma$ and $\gamma$ for the metastable and the excited states,
respectively, of the atom.

%

The full widths at half maximum are
\begin{subequations}
\begin{align}
F_{1} & =\frac{\Gamma+\gamma}{2}-(\Gamma-\gamma)\zeta,  \label{3-12-1} \\
F_{2} & =\frac{\Gamma+\gamma}{2}+(\Gamma-\gamma)\zeta,  \label{3-12-2}
\end{align}
which is derived from the first order Taylor's expansion by assuming large
quality factors $\Gamma/\omega_{\mathrm{e}}$ and $\gamma/\Delta.$

It is obvious that the scattering of the photon by the $\Lambda$-type
three-level atomic FQN can be regarded as scattering by two two-level FQN's
at the same position. This fact can be seen from the detailed calculation
about the transmission and reflection by two-level FQN's in Ref.~\cite{zhou}%
. The main conclusion is that the FQN play the same role as that of a $%
\delta $-potential in the 1D scattering problem. Tuning the FQN properly can
establish an infinite potential barrier to totally reflect the incident
single photon. The detailed discussions are given in the next section.

\section{\label{sec:reflection}CONTROLLABLE REFLECTION AND TRANSMISSION}

\subsection{The tunable double peak}

For the coherent transport of a single photon in the 1D continuum, the
scattering equation
\end{subequations}
\begin{equation}
(E-\omega)u(j)=-tu(j+1)-tu(j-1)  \label{4-1}
\end{equation}
for $j\neq0$ assumes a usual solution%
\begin{equation}
u(j)=\left\{
\begin{array}{c}
e^{-ikj}+re^{ikj},j<0 \\
se^{-ikj},\text{ \ \ \ \ \ }j>0%
\end{array}%
\right.,  \label{4-2}
\end{equation}
where $r$ and $s$ are the reflection and the transmission coefficients,
respectively. The cavity lattice constant is normalized to 1. Apparently,
the energy $E$ of the incident photon obeys the dispersion relation
\begin{equation}
E=\omega-2t\cos(k),
\end{equation}
dependent on the momentum $k$ the of incident photon.

The continuous condition $u(0^{+})=u(0^{-})$ together with scattering
equation at the zeroth site
\begin{equation}
[E-\omega-V]u(0)=-tu(1)-tu(-1)  \label{4-3}
\end{equation}
determines the reflection coefficient
\begin{equation}
r=\frac{V}{2it\sin(kl)-V},  \label{4-4}
\end{equation}
and the transmission coefficient $s=1+r.$

%
\begin{figure}[ptb]
\begin{centering}
\includegraphics[bb=25bp 225bp 580bp 689bp,clip,width=3.5in]{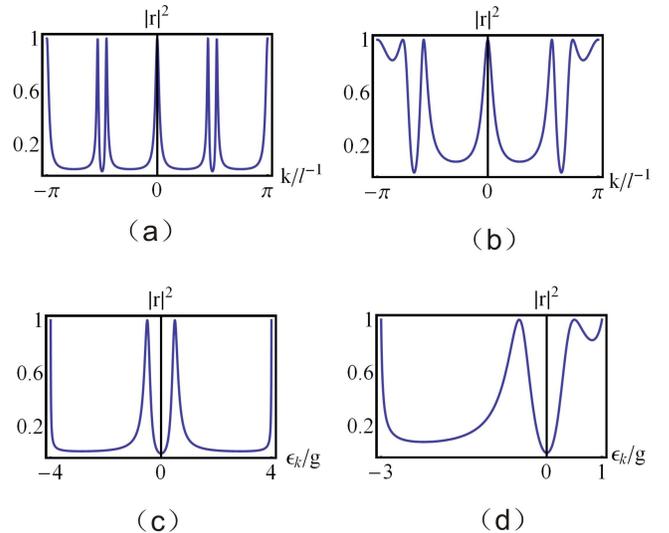}
\par\end{centering}
\caption{Reflection coefficients vs. $k$ and $\protect\varepsilon_{k}$. (a)
and (b) is plotted versus momentum $k$ of incident photon, and (c) and (d)
is plotted versus energy $\protect\varepsilon_{k}$. Here the parameters in
our setup is chosen as: (1)$t=2,$ $\protect\omega=1,$ $\protect\omega_{%
\mathrm{e}}=1,$ $\Delta=0,$ $\Omega=1$ (all is in units of coupling constant
$g$) for (a) and (c); (2)$t=1,$ $\protect\omega=1,$ $\protect\omega_{\mathrm{%
e}}=2,$ $\Delta=0,$ $\Omega=0.75$ (all is in units of coupling constant $g$)
for (b) and (d).}
\label{fig:lineshape}
\end{figure}


The reflection coefficient for single photon transport is plotted in Fig.~%
\ref{fig:lineshape}. The reflection coefficient is plotted against the
momentum $k$ of the incident photon in Fig.~\ref{fig:lineshape}(a) and Fig.~%
\ref{fig:lineshape}(b), and against the energy $\varepsilon_{k}=E(k)-\Delta$
in Fig.~\ref{fig:lineshape}(c) and Fig.~\ref{fig:lineshape}(d). Obviously,
there are rich lineshapes beyond the conventional Breit-Wigner and
Fano-Feshbach types.

For a single incident photon with a definite momentum, the potential $V(j)$
determines all the properties of a scattering process. The potential is only
located at the zeroth site where the $\Lambda$-type atom is placed. The
nontrivial cases%
\begin{equation}
r=\left\{
\begin{array}{c}
0,\text{ perfect transmission} \\
-1,\text{ perfect reflection}%
\end{array}%
\right.
\end{equation}
occurs when the potential $V(j)$ takes special values. The zeros of $V(j)$
correspond to perfect transmission, and the singularities of $V(j)$
correspond to perfect reflection. Therefore, we can control reflection and
transmission by tuning the Rabi frequency $\Omega$ and the control field
frequency $\omega_{\mathrm{C}}$. Similar to the phenomenon of the negative
differential electric resistance in an electronic transistor, we observe
here a {}``negative differential photonic resistance\textquotedblright.
Following this comparison, we have demonstrated an all-optical device.

From Eq.(\ref{3-6}), the energy $E$ of the single photon at perfect
transmission, i.e. $V=0$, satisfies%
\begin{equation}
E-\Delta =0,  \label{4-6}
\end{equation}%
which is exactly the two-photon resonant condition. The photons scattered
from the two potential peaks interfere coherently such that the back
traveling photon is eliminated while the forward traveling photon is
constructed, which gives perfect transmission to the incident photon. This
phenomenon never occurs in 1D scattering problem with a two-level FQN in
Ref.~\cite{zhou} because of the lack of the EIT mechanism in a two-level
atom. The case of ideal transparency can be easily found in Fig.~\ref%
{fig:lineshape}, where in the vicinity of $\varepsilon _{k}=0$ the original
single peak in Ref.~\cite{zhou} splits into two peaks. The splitting
position can be tuned by varying the Rabi frequency of the classical control
field.

The other nontrivial case is the perfect reflection, where the three-level
atom acts like a perfect {}``mirror\textquotedblright. The perfect
reflection caused by the singularities of the potential $V$ takes place only
when the energy of the incident photon resonates with one of the internal
level spacings of the three-level atom.

%
\begin{figure}[ptb]
\begin{centering}
\includegraphics[bb=82bp 341bp 506bp 671bp,clip,width=3in]{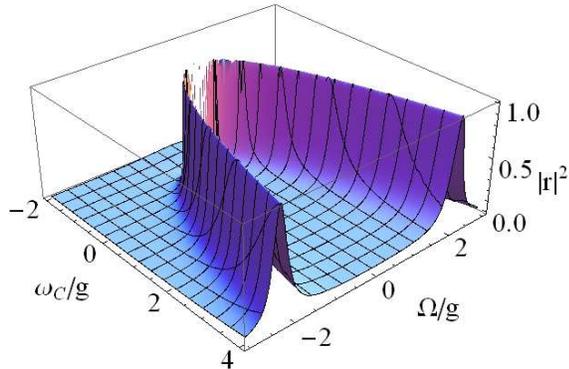}
\par\end{centering}
\caption{(Color online) 3D plot of reflection coefficients vs Rabi frequency
$\Omega$ and energy level spacing of classical light field $\protect\omega_{%
\mathrm{C}}$ (Both $\Omega$ and $\protect\omega_{\mathrm{C}}$ is in units of
coupling constant $g$).}
\label{fig:reflection}
\end{figure}


In Fig.~\ref{fig:reflection}, we can see how the reflection coefficient
varies with the Rabi frequency $\Omega$, as the x-axis, and with the
frequency $\omega_{\mathrm{C}}$ of the external control field, as the
y-axis. The existence of the double peaks in the plot shows that when fixing
one of the parameters $\Omega$ or $\omega_{C}$, we can always fine tune the
other parameter to reach perfect reflection or transmission. The reason why
there always exist two peaks for perfect reflection is that the $\Omega^{2}$
term in the reflection coefficient corresponds to two values, $\Omega$ and $%
-\Omega$, of the Rabi frequency.

\begin{figure}[tbp]
\begin{centering}
\includegraphics[clip,width=3.5in]{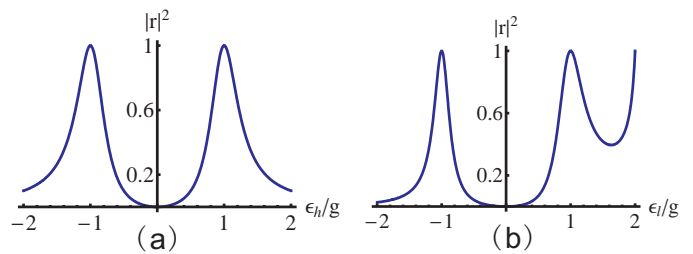}
\par\end{centering}
\caption{The reflection coefficients as functions of $\protect\varepsilon_{%
\mathrm{m}}$ are plotted under (a) the high energy limit and (b) the low
energy limit. None of the lineshapes are simple Breit-Wigner and
Fano-Feshbach types.}
\label{fig:limit}
\end{figure}

The high and the low energy limits are obtained in the vicinity of $k=\pi/2$
and $0$, respectively. The reflection coefficients become ($h$ and $l$ index
the high and the low energy limits, respectively)%
\begin{equation}
r=\left\{
\begin{array}{c}
V_{\mathrm{h}}/(2it-V_{\mathrm{h}}),(k\rightarrow\pi/2) \\
V_{\mathrm{l}}/(2itk-V_{\mathrm{l}}),(k\rightarrow0)%
\end{array}%
\right.,  \label{4-10}
\end{equation}
where the corresponding potentials in the high and the low energy limits
equal to
\begin{equation}
V_{\mathrm{m}}=\frac{g^{2}\varepsilon_{\mathrm{m}}}{(\Delta\varepsilon_{%
\mathrm{m}}+\Delta-\omega_{\mathrm{e}})\varepsilon_{\mathrm{m}}-\Omega^{2}}%
,(m=h,l).  \label{4-11}
\end{equation}
The reflection coefficient is plotted against the energy $\varepsilon_{%
\mathrm{m}}$ in Fig.~\ref{fig:limit}: (a) in the high energy limit as a
function of $\varepsilon_{\mathrm{h}}$ where the energy obeys the linear
dispersion relation $\varepsilon_{\mathrm{h}}=(\omega-t\pi-\Delta)+2tk$, and
(b) in the low energy limit as a function of $\varepsilon_{\mathrm{l}}$
where the energy obeys the quadratic dispersion relation $\varepsilon_{%
\mathrm{l}}=(\omega-2t-\Delta)+tk^{2}$. Obviously, the lineshapes are
different from the conventional Breit-Wigner and Fano-Feshbach types.

\subsection{Equivalence to a pair of two-level atoms}

To show explicitly that the three-level FQN is equivalent to a pair of
two-level atoms placed apart in the cavity-array, we consider the scattering
problem in the 1D continuum with one two-level atom deposited at the zeroth
site and the other at the $D$-th site. By a similar approach in discrete
coordinate representation used in Ref.{[}zhou], we obtain the reflection and
the transmission coefficients
\begin{subequations}
\begin{align}
r & =\frac{V_{1}f_{2}(k)\exp(i2kD)-V_{1}V_{2}+V_{2}2ti\sin k}{%
-f_{1}(k)f_{2}(k)\exp(i2kD)+V_{1}V_{2}},  \label{4-7-1} \\
s & =\frac{(2t\sin k)^{2}\exp(i2kD)}{-f_{1}(k)f_{2}(k)\exp(i2kD)+V_{1}V_{2}},
\label{4-7-2}
\end{align}
where the transport functions are defined
\end{subequations}
\begin{equation}
f_{\mathrm{m}}(k)=2si\sin k+V_{\mathrm{m}},(m=1,2)  \label{4-8}
\end{equation}
with the potentials
\begin{equation}
V_{\mathrm{m}}=\frac{g_{\mathrm{m}}^{2}}{E-\omega_{\mathrm{m}}},(m=1,2).
\label{4-9}
\end{equation}

%

The reflection and the transmission coefficients in Eqs.(\ref{4-7-1}) and (%
\ref{4-7-2}) are identical to the ones in our setup, besides the additional
phase factor $\exp(i2kD)$ determined by the momentum $k$ of the incident
photon and the distance $D$ between the two FQN's. We observe that the
effect of $D$ is totally contained in this phase factor, which equals to the
phase difference between the incident wave and the reflected wave from the $%
D $-th site. The two effective potentials in Eq.(\ref{4-9}) together play
the same role as the complex potential in Eq.(\ref{3-6}). If we treat the
two atoms with the cavities in between as an extended
{}``FQN\textquotedblright, then for a certain momentum $k$ of incident
photon, the phase factor becomes $\exp(i2kD)=1$ and this extended
{}``FQN\textquotedblright is equivalent to a $\Lambda$-type atom in our
setup.

\begin{figure}[ptb]
\begin{centering}
\includegraphics[bb=54bp 368bp 543bp 577bp,clip,width=3.5in]{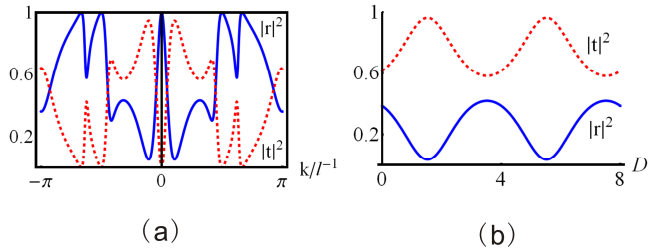}
\par\end{centering}
\caption{(Color online) (a) The reflection and the transmission coefficients
vs. $k$ for two distant two-level FQN's. (b) The reflection and the
transmission coefficients vs. $D$ for two distant two-level FQNs. Blue solid
and red dashed line represent reflection and transmission coefficients,
respectively.}
\label{fig:2a2}
\end{figure}

For a fixed distance $D$, the transmission and the reflection coefficients
are plotted in Fig.~\ref{fig:2a2}(a). In this setup, the perfect reflection
can be attained when both the effective potentials $V_{1}$ and $V_{2}$ tend
to infinity. However, the existence of perfect transmission is not
guaranteed. Not only the two-photon resonance condition but also the
distance constrain $\exp(i2kD)=1$ must be met to attain perfect
transmission. For a fixed momentum $k$, the periodic variance of the
transmission and the reflection coefficients is shown in Fig.~\ref{fig:2a2}%
(b).

%

%
\begin{figure}[tbp]
\begin{centering}
\includegraphics[bb=0bp 576bp 595bp 809bp,clip,width=3.5in]{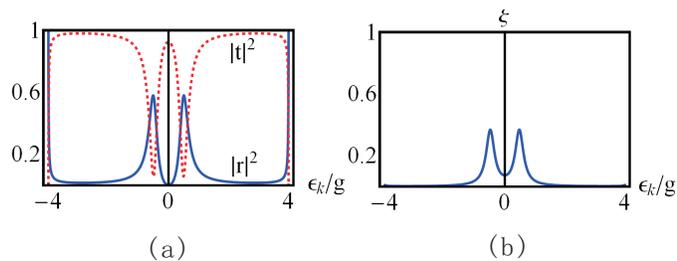}
\par\end{centering}
\caption{Reflection coefficients and the energy loss ratio vs. $\protect%
\varepsilon_{k}$,respectively in (a) and (b). Here the parameters in our
setup is chosen as: $t=2,$ $\protect\omega=1,$ $\protect\omega_{\mathrm{e}%
}=1,$ $\Delta=0,$ $\Omega=1$, $\protect\gamma=\Gamma=0.04$ for both (a) and
(b) (all is in units of coupling constant $g$)(all is in units of coupling
constant $g$) for (b).}
\label{fig:decay}
\end{figure}


\subsection{Cavity decay and atomic decay}

The material and devices imperfections result in unadvoidable energy
relaxation and dephasing of devices. Such a decoherence effect results in
the inelastic scattering of a single photon and reduce the switching
efficiency.

For atomic decay, it can be simply demonstrated by phenomenologically
introducing two additional atomic decay rates $\Gamma $ and $\gamma $ for
the metastable and the excited states, respectively, of the atom. According
to the Eqs. (\ref{3-12-1}) and (\ref{3-12-2}), the maximum value of
effective potential is decreased to a finite value, which implies that
perfect reflection would not be obtained any more.

To derive the scattering property of the propagation of a single photon with
atomic decays $\Gamma $ and $\gamma $, the frequencies $\omega _{\mathrm{e}}$
and $\Delta $ are substituted by $\omega _{\mathrm{e}}-i\Gamma $ and $\Delta
-i\gamma $ to phenomenologically represent the atomic energy relaxation. The
straightforward calculation gives the reflection amplitude as
\begin{equation}
r=\frac{V_{d}}{2it\sin (kl)-V_{d}},
\end{equation}%
where
\begin{equation}
V_{d}=\frac{g^{2}(E-\Delta +i\gamma )}{(E-\omega _{\mathrm{e}}+i\Gamma
)(E-\Delta +i\gamma )-\Omega ^{2}}
\end{equation}%
and the corresponding transmission amplitude $s=1+r.$ The reflection
coefficient for single photon transport with atomic decay is plotted in Fig.~%
\ref{fig:decay} (a). \ \

Obviously, the perfect transparency can not be obtained even when the
absolute value of effective potential is zero. The maximun value of
transmission coefficient is less than $1$ depicted in Fig.~\ref{fig:decay}(a)%
$.$ At the meantime, the maximum value of reflection coefficient depicted as
peaks in Fig.~\ref{fig:decay}(a) are dramatically decreased. And the
summation of reflection coefficient and transmission coefficient is always
less than $1,$ which implys that the single photon undergo an inelastic
scattering process. This inelastic effect is plotted in Fig.~\ref{fig:decay}
(b). The label of y-axis $\xi =1-(\left\vert r\right\vert ^{2}+\left\vert
t\right\vert ^{2})$ represents the ratio of loss energy after the inelastic
scattering.

To depict the non-ideal scattering process with various decoherences, we
need to consider the coherent length (CL) of the scattering process, and an
infinite CL means an ideal elastic scattering. Here, the CL can be regarded
as the distance that the photon travels between the left side and the right
side of the scatterer (the three-level system). Actually, to make sure that
the scattering process can happen, it is required that the photon leakage
rate $\kappa $ for each resonator is much smaller than the hopping constant,
otherwise photons will totally escape into the environment before
encountering the scatterer. Therefore, we think that the photon leakage rate
$\kappa $ defines the CL, which is roughly proportional to the product of $%
\kappa ^{-1}$ and the group velocity of the photons. Moreover, any
additional change of the leakage rate, at the point where the scatterer is
located, will broaden the width of the lineshape at the resonance (i.e.,
there is a peak at the transition energy).

To study in detail, this decoherence effect, we need to use a microscopic
model where both the cavities and the there-level system are coupled to the
external environment. The decoherence of every cavity and the there-level
system mainly results in the incoherent or dissipative propagation of the
incident photon. In such approach, the decoherence or dissipation can be
divided into two categories according to its contributions to the scattering
process. One category contains all the cavity decays, except the one of the
cavity which is directly coupled to the there-level system, since these
resonators contribute to the free propagation of the photon. This trivial
type of photon leakage rate in each cavity only affects the coherent length
of the scattering process. The other category influencing the scattering
process contains the decay of the there-level system and the decay of the
cavity directly coupled to the two-level system. Because the energy of the
incident photon is not conservative before and after the scattering happens,
the scattering is obviously inelastic. The inelastic scattering process
would broaden the width of the lineshape. These different decoherence
categories would reduce the quantum switching efficiencies in different ways.

\section{\label{sec:quasibound}FORMING A SECONDARY CAVITY WITH TWO FQN'S}

According to the results derived above, we now propose a controllable
secondary cavity architecture. We place two $\Lambda$-type FQN's at the
zeroth site and the $D$-th site in the 1D cavity-array (Fig.~\ref{fig:model}%
(c)). These two FQN's acting as $\delta$-potentials with tunable potential
are controlled by external classical light fields. The most natural
consideration is that there exists a quasibound state between the two $%
\delta $-potentials, which can degenerate to become a bound state under some
special cases. In our setup, it is intuitive that the two $\Lambda$-type
FQN's form a secondary cavity for single photon storage among the regular
cavities in the array. We can prove that only incident photons of some
particular momenta can be stored in this secondary cavity.

The total Hamiltonian in the {}\textquotedblleft rotating\textquotedblright
frame of reference

\begin{equation}
H^{\mathrm{R}}=H_{\mathrm{p}}+H_{\mathrm{a}}^{\prime }+H_{\mathrm{c}%
}^{\prime }  \label{5--1}
\end{equation}%
reads
\begin{eqnarray}
H_{\mathrm{p}} &=&\underset{j}{\sum }[\omega b_{j}^{\dag
}b_{j}-t(b_{j}^{\dag }b_{j+1}+h.c)],  \label{5-2} \\
H_{\mathrm{a}}^{\prime } &=&\underset{l=1,2}{\sum }[\omega _{\mathrm{e},%
\mathrm{l}}\left\vert e\right\rangle \left\langle e\right\vert _{\mathrm{l}%
}+\Delta _{\mathrm{l}}\left\vert a\right\rangle \left\langle a\right\vert _{%
\mathrm{l}}],  \label{5-3} \\
H_{\mathrm{c}}^{\prime } &=&\underset{\mathrm{l}=1,2}{\sum }[\Omega _{%
\mathrm{l}}(\left\vert a\right\rangle \left\langle e\right\vert _{\mathrm{l}%
}+h.c.)+g_{\mathrm{l}}(b_{\mathrm{l}}^{\dag }\left\vert g\right\rangle
\left\langle e\right\vert _{\mathrm{l}}+h.c.)],  \label{5-4}
\end{eqnarray}%
where $H_{\mathrm{p}}$ describes the free energy of the photon, $H_{\mathrm{a%
}}^{\prime }$ the potential energy of the two FQN's with detunings $\Delta _{%
\mathrm{l}}=\omega _{\mathrm{a},\mathrm{l}}-\omega _{\mathrm{C},\mathrm{l}}$
($l=1$ indicates the FQN at the zeroth sites while $l=2$ that of the $D$-th
site), and $H_{\mathrm{c}}^{\prime }$ the interaction of the FQN's with the
photon probe and the classical control fields.

Again, the energy eigenvectors can be expanded in the basis of an invariant
subspace in the form%
\begin{equation}
\left\vert E\right\rangle =\underset{j}{\sum}u(j)b_{j}^{\dag}\left\vert
0,g,g\right\rangle +\underset{\mathrm{l}=1,2}{\sum}[u_{\mathrm{a}}^{\mathrm{l%
}}\left\vert 0,a,g\right\rangle +u_{\mathrm{e}}^{\mathrm{l}}\left\vert
0,e,g\right\rangle ],  \label{5-5}
\end{equation}
where $u(j)$ is the probability amplitude of the single photon at the $j$-th
site. Following the same procedure as in the one FQN case, we obtain the
scattering equation of the probability amplitudes
\begin{equation}
[E-\omega-\underset{\mathrm{l}=1,2}{\sum}V_{\mathrm{l}}\delta_{j,\mathrm{l}%
}]u(j)=-tu(j-1)-tu(j+1),  \label{5-6}
\end{equation}
where the potentials at the two sites of FQN's are
\begin{equation}
V_{\mathrm{l}}=\frac{g_{\mathrm{l}}^{2}(E-\Delta_{\mathrm{l}})}{(E-\omega_{%
\mathrm{e},\mathrm{l}})(E-\Delta_{\mathrm{l}})-(\Omega_{\mathrm{l}})^{2}}%
,(l=1,2).  \label{5-7}
\end{equation}
It should be noted that both the potentials act like $\delta$-potentials,
between which a quasibound state can survive.

The quasi-bound state can be considered the analytic continuation of a
scattering state into the complex momentum plane singular reflection and
transmission coefficients. Accordingly, we first assume the scattering state
to be%
\begin{equation}
u(j)=\left\{
\begin{array}{c}
e^{-ikjl}+re^{ikjl},\text{ \ \ \ \ }j<0 \\
s_{1}e^{-ikjl}+r_{1}e^{ikjl},0<j<D, \\
se^{-ikjl},\text{ \ \ \ \ \ \ \ \ \ \ }j>D%
\end{array}%
\right..  \label{5-8}
\end{equation}
where $r$ and $s_{1}$ are the reflection and the transmission coefficients
at the zeroth site, whereas $r_{1}$ and $s$ are those at the $D$-th site.
Since we are only interested in the effective behavior of the secondary
cavity as an intact storage device, the intra-cavity analysis of wave
transmission can be neglected. That is, we are only concerned about the
coefficients $r$ and $s$ viewed from outside the secondary cavity, whose
values are the solution of the boundary value problem Eq.(\ref{5-8}) and
read
\begin{subequations}
\begin{align}
r & =\frac{V_{1}f_{2}(k)\exp(i2kD)-V_{1}V_{2}+V_{2}2ti\sin k}{%
-f_{1}(k)f_{2}(k)\exp(i2kD)+V_{1}V_{2}}, \\
s & =\frac{(2t\sin k)^{2}\exp(i2kD)}{-f_{1}(k)f_{2}(k)\exp(i2kD)+V_{1}V_{2}}.
\end{align}
Except that the potentials $V_{1}$ and $V_{2}$ are defined as in Eq.(\ref%
{5-7}), the above expressions are identical to Eq.(\ref{4-7-1}) and Eq.(\ref%
{4-7-2}), which leads to perfect transmission through the secondary cavity
based on EIT mechanism.

The quasi-bound state then occurs when the denominators of the two
coefficients equal to zero, which corresponds to the condition
\end{subequations}
\begin{equation}
e^{i2kD}=\frac{V_{1}V_{2}}{(2ti\sin k+V_{1})(2ti\sin k+V_{2})},  \label{5-9}
\end{equation}
from which the momentum $k$ of a single photon surviving between the
potentials $V_{1}$ and $V_{2}$ is determined. The incident photon
corresponding to this intra-cavity photon has its momentum take imaginary
values, which in turn leads to its imaginary energy. The imaginary energy
will result in a decay of the wavefunction, which means the quasi-bound
state is a bound state with a small leakage at two ends. When $\omega\gg t$,
it is almost impossible for a single photon to propagate in the channel and
quasi-bound states are formed in the secondary cavity.

The most interesting quasi-bound state can be obtained when both FQN's are
tuned to their resonant states, where%
\begin{equation}
e^{-i2kDl}=1.  \label{5-10}
\end{equation}
The above formula subjects the momentum of the trapped photon to a quantized
value
\begin{equation}
k=\frac{\pi n}{Dl},  \label{5-11}
\end{equation}
where $n$ is an arbitrary integer. The corresponding bound state is%
\begin{equation}
u(j)=\left\{
\begin{array}{c}
0,\text{ \ \ \ \ }j<0\text{ or }j>D \\
A\sin(kj),0<j<D%
\end{array}%
\right.  \label{5-12}
\end{equation}
where $A$ is the normalization constant. So the incident photon can be
perfectly trapped between the two FQN's, similar to what happens in a single
cavity. We hence name this setup a perfect secondary cavity, which realizes
a non-destructive single photon storage.

Compared to the proposals of quantum memories based on the dark
state in the EIT effect, our secondary cavity setup is much easier
to implement. In the former, the dark state is stationary and the
classical field is adiabatically manipulated to store information
from the incident photon to the three-level atom. Such a process
should be regarded as a stationary storage and demands highly
precise control over the external control field. Whereas, the
secondary cavity proposed here is dynamic and the control
parameters, including the distance $D$ between the FQN's, the Rabi
frequency $\Omega$, and the frequency $\omega_{\mathrm{C}}$ of the
external field, are all much easier to manipulate. The storage
process can be imagined as disposing two perfectly reflecting
``mirrors'' in the cavity-array such that a single photon originally
traveling in the array is bounded in between to become a standing
wave. The releasing process is to inversely remove these two
reflecting ``mirrors''.

\section{\label{sec:conclusion}CONCLUSION}

We have revisited the problem of single photon transport in an 1D
cavity-array with a deposited three-level $\Lambda $-type FQN and
illustrated how the reflection and the transmission coefficients rely on the
Rabi frequency and the traveling frequency of a classical control field
external to the cavity array. By tuning these two frequencies, the FQN can
serve as a perfect {}\textquotedblleft mirror\textquotedblright\ or a
transparent medium for an incident photon. The appearance of lineshapes
different from the conventional Breit-Wigner or Fano-Feshbach type was shown
to stem from a nonlinear dispersion relation and the EIT mechanism in the
setup.

The dissipation of cavities and atom are taken into account. The atomic
decay is added phenomenologically to explain the single photon undergo an
inelastic scattering process. The cavities decay mainly determine the
coherent length, which limit the upper number of cavities.

Using this phenomenon, we have proposed a secondary cavity between two
FQN's, in which a controllable quasibound state can be formed, to coherently
store a single photon. A perfect secondary cavity selects photons with
momenta being integral mutiples of a constant. The limiting lossless case
has also been presented to compare with the usual photon storage using an
EIT dark state.

\begin{acknowledgments}
The authors thank H. Dong, T. Shi and D. Z. Xu for helpful discussion. This
work is supported by the NSFC with Grants No.90203018, No.10474104,
No.60433050 and No. 10704023. It is also funded by the National Fundamental
Research Program of China with Grants No.2001CB309310 and No.2005CB724508.
\end{acknowledgments}

\end{document}